# PSYCHIATRIC HOME TREATMENT FOR INPATIENT CARE CASES – DESIGN, IMPLEMENTATION AND PARTICIPATION

Stefan Hochwarter, Department of Computer Science, Norwegian University of Science and Technology, Trondheim, Norway

Pierre Tangermann, IZI-BB, Fraunhofer-Institute for Cell Therapy and Immunology, Branch Bioanalytics and Bioengineering

Martin Heinze, Brandenburg Medical School, Clinic for Psychiatry and Psychotherapy, Immanuel Clinic Rüdersdorf

Julian Schwarz, Brandenburg Medical School, Clinic for Psychiatry and Psychotherapy, Immanuel Clinic Rüdersdorf

## ABSTRACT

*The use of information and communication technologies (ICT) to support long-term care is gaining attention, also in the light of population ageing. Known in Scandinavian countries under the term of welfare technology, it aims to increase the quality of life and independence of people with physical, psychological or social impairments. In Germany, a new form of psychiatric home treatment, inpatient equivalent treatment (IET), is offered since 2018. It should allow service users with severe mental health issues to stay in their familiar environment during crisis, while being treated in the same complexity and flexibility like in an inpatient unit. However, this change in delivering healthcare services leads to socio-technical challenges, such as coordination of work, integration into existing healthcare workflows and ensuring continuity of care. Hence, the objective of this exploratory study is to examine how information and communication technologies (ICT) interact in the new setting and how this process can be improved. Further, we also ask how service users can participate in designing home treatment services.*

*Methodologically, this study follows a qualitative research approach. Different methods including participant observation, interviews and focus groups were conducted to answer the research questions. Data was collected during a field visit at the psychiatric department of a German clinic in summer 2019. Field notes and interviews were analyzed using the R package for qualitative data analysis RQDA. A list of socio-technical challenges and opportunities related to IET were identified. New forms of communication, gaps in documentation practices and continuity of care are seen to be highly relevant for designing and implementing home treatment services in psychiatric care. We also discuss how service users and health professionals can take pro-active part in designing these services.*

## 1 INTRODUCTION

It is well known that population projections indicate a worldwide population ageing with high-income countries leading the list. Japan has the world's most aged population, followed by Germany (UN, 2017). In Norway, within the next 15 years the population of children and young people will be overtaken by the elderly population (Syse et al., 2018). While this demographic shift certainly poses a challenge to various healthcare services, the OECD also stresses the economic component of rapid ageing and urges to the countries to rethink their long-term care policies (Swartz, 2013).

In this light, welfare technology is gaining a lot of attention recently. The term "welfare technology" origins from Denmark and is established in Scandinavian countries. Similar terms in English would include e-health, telecare or ambient assisted living. They share their common aim to strengthen its users' independence and support to master their everyday life on their own (Helsedirektoratet, 2012). Popular examples of welfare technology focus on the treatment and management of chronic conditions such as chronic obstructive pulmonary disease (COPD) or diabetes. However, when looking at the definition of welfare technology, it explicitly includes people with psychosocial impairments. Further, the definition also includes services that support access and improve efficiency of services (Departementenes servicesenter, Informasjonsforvaltning, 2011). In this paper we focus on the delivery of psychiatric home treatment, a service were service users (patients) are treated outside the clinic and can remain in their familiar environment, normally their homes. We investigate how existing information and communication technology (ICT) infrastructure promotes or hinders the delivery of home treatment.

This paper is building on the findings of a summer school organised by the European Institute for Innovation and Technology (EIT). The aim of the summer school was to "*train attendees on the subject of technological innovation in health by giving a global vision of a medical device maturation cycle, from the idea to the market*" ('ClinMeD2019', 2009). The process to transform medical ideas into a commercialised product, device or service is highly complex and heterogeneous. Identifying needs by observation and interaction with end-users (e.g. healthcare professionals, service users and advocacy groups) can contribute to a well adapted design of new medical devices by providing not only relevant technical specifications but also user-oriented system requirements. To this end, the presented study was performed within the scope of ClinMed19, an European Union funded education and research action based on these participatory concepts and, therefore, designed to transport medical innovations into market applications.

We draw mainly on theories in the fields of Computer Supported Cooperative Work (CSCW), as we believe the academic field of CSCW fits well to investigate the underlying case. While our empirical case is specific, we believe our findings contribute to the design of similar welfare technology systems. This study addresses the following research questions: (i) *How is psychiatric home treatment carried out?* (ii) *How is psychiatric home treatment documented, coordinated and integrated into the clinical workflows?* (iii) *How can we design psychiatric home treatment systems that support patient participation?*

The first research question aims to understand the work that involves carrying out psychiatric home treatment. A clear picture of the phenomena at hand informs the two other research questions. While the first research question allows a rather wide exploration, the second and third research questions go into specifics. Research question two investigates the use of ICT systems for the work described in the first research question. The last research question aims to shift the perspectives from the care providers towards the service users and how they can be possibly included in this process.

The rest of this paper is structured as follows: first, we present the background and introduce the context. Second, we describe the methods used for data collection and analysis. Next, we present our results and continue with a discussion addressing the stated research questions. Finally, we sum up our study with a conclusion, limitations and future work.

## 2 BACKGROUND

We are interested in how psychiatric home treatment is carried out with the help of ICT, especially in regards to documentation, coordination and integration. Further, we also want to throw light on patient participation, as this is one central value of the psychiatric department at hand. Certainly these topics are not new, hence, we will first summarise related work and then move to the political and organizational context of this study.

### 2.1 Related work

ICT in healthcare has been studied from different perspectives in different areas. The use of ICT is often propagated to deliver healthcare services more efficient and effective. This is not always the case. There are many challenges and pitfalls when designing and implementing such systems. In CSCW, where the focus is on understanding cooperative work to design ICT artefacts that support this work, the area of healthcare has been and still is a major field of studies. It is not surprising, that a lot of relevant findings, also for home treatment, are from this area. CSCW underlines that "*healthcare work is highly institutionalised and complex, involves multiple stakeholders, takes place across primary, secondary and tertiary care sectors, in private or public funding arrangements, and depends on a highly collaborative approach*" (Fitzpatrick & Ellingsen, 2013). Now the context of work in healthcare also moves into the home of the service users, hence making this field even more complex. Coordination of work that includes direct actors (such as service users) and indirect actors (such as informal caregivers) is challenging for the design of ICT systems. Also the design of documentation systems for this specific context change the requirements for interoperability and integration (Fitzpatrick & Ellingsen, 2013).

Interestingly, there are a few concepts on that everyone seems to agree that they are obviously beneficially. (Tight) integration of systems is one of these. It is worth noting that alternative strategies exist and it should be evaluated to which degree integration is in the given context actually beneficially (Monteiro, 2003). Another strategy people tend to support without hesitation is the elimination of redundancy, which is often seen as harmful or waste. A recent study pointed out the advantages of having certain redundancies and that they can actually increase safety and carry informal information that would otherwise be lost (Cabitza & Simone, 2013).

The clinic follows a "participatory approach", as they like to stress. Hence, we included a research question asking how we can design a system for psychiatric home treatment that are in line with this participatory approach. Our paper draws on concepts and practices from CSCW, a field that is closely connected to the Participatory Design community. Participatory Design (PD) has a strong political and social commitment: "*Perhaps the core principle of participatory design is that people have a basic right to make decisions about how they do their work and indeed any other activities where they might use technology.*" (Simonsen & Robertson, 2012) PD aims to include the users in all design processes and decision of a project. Done right, this is a very difficult and time-consuming task. This was also pointed out by (Batalden et al., 2016) when designing healthcare services in co-production with service users (although they don't reference PD, they follow very similar principles). These mentioned challenges, besides others, are currently reason for a lively discussion in the PD community (Wagner, 2018, p. 201). Although, this paper will not contribute to this discussion, we believe it is important for this study to know that fundamental discussions are taking place.

## 2.2 Political context

Home-Treatment (HT) as alternative for psychiatric hospital treatment has initially been introduced in the 1980s in the US, followed by Great Britain, the Netherlands and Scandinavian countries (Stein & Test, 1980). Since then this form of treatment has been further developed and differentiated to several sub-forms, addressing service users with different needs of psychiatric support. Most contemporary approaches to HT are *Crisis Resolution and Home Treatment Teams* (CRHTs), offering a treatment of acute psychiatric crisis in the service users' home provided by multi-professional outreach teams, available 24h/7 days a week. Current evidence shows, that CRHT is an efficient and effective model of treatment, which is in several aspects superior to traditional inpatient treatment (Murphy et al., 2015). CRHTs may reduce the number of inpatient stays and lower the overall treatment costs (Johnson et al., 2005; McCrone et al., 2009), improve psychosocial functioning and symptom severity, while service users' have made predominantly positive experiences (Winness et al., 2010).

In contrast to that, Germany can be regarded as a late adopter of intensive psychiatric outreach services. One major reason for that can be seen in the fragmentation of the German health and the social system into different social codes, strictly separating inpatient and outpatient sectors (H.J. Salize & M. Lambert, 2016). This has resulted to a multitudinous range of services and service providers in communal psychiatry, hampering cooperation, need-adaptation and the introduction of modern psychiatric services (Schwarz et al., 2019a). To overcome these issues in 2013 a new legislation (§64b Social code V) was introduced, promoting the implementation of integrated models of care, offering a continuous treatment across different sectors, including outreach home care. First evaluation results indicate positive outcomes both for services users, informal caregivers and professionals (Schwarz et al., 2019b; von Peter et al., 2019).

To further develop mental health services and based on positive experiences with the integrated models of care, finally in 2018 a legislation was adopted, introducing Inpatient Equivalent Treatment (IET) as a form of HT in standard psychiatric care (according to §115d Social Code V). It was defined as a multi-professional, team-based outreach approach for people suffering from an acute psychiatric crisis. Contrary to CRHT, IET requires from the offering hospital to visit the service user daily at his home environment, likewise there is no 24/7- outreach support. Since 2018, only a few hospitals have introduced IET yet, which is due to the high (infra-)structural requirements, as well as the partly unsecured financing of the service (Längle et al., 2018).

### 2.3 Organisational context

Focusing on our research field, the university clinic for psychiatry and psychotherapy of Brandenburg Medical School ("Immanuel Clinic") is situated in Rüdersdorf, Brandenburg. It ensures acute psychiatric services for a catchment area encompassing approx. 245.000 inhabitants. Located in the outskirts of Berlin, it includes both sub-urban and very rural areas. While Immanuel Clinic Rüdersdorf (ICR) had already implemented Integrated Care in 2016, it was one of the first hospitals in Germany having introduced IET in May 2018. Since then it is running one IET-team, equipped with altogether 7 employees and two cars, covering major parts of the catchment area.

Besides the experiences with innovative treatments elements, ICR has a long tradition in involving service users ("peer support workers") in their services, service development and research projects ("psychiatric survivor researchers"), currently holding one of the few chairs for participatory and collaborative research in Germany (von Peter, 2017). One of the key ideas of the service user involvement in psychiatric practise and research in ICR is to make decision processes more democratic in relation to both treated and treating stakeholders and in doing so, to achieve a generally higher need adaptation of the services.

## 3 METHODS

As stated previously, this study builds on findings made during a summer school from 02.07.2019 to 12.07.2019, consisting of two parts: (i) an immersive field experience at the study site in Rüdersdorf, Brandenburg, Germany; and (ii) the design phase at the main site in Autrans, France.

The results described in this paper rely heavily on the data collected during our field visit at the Immanuel Clinic Rüdersdorf, still we consider our gained insights and discussions in Autrans of being constructive and relevant to the overall picture we are presenting here. Hence, this section is divided into two parts. First we will describe the overall methods and frameworks used during the summer school, and second we focus on the methods and data collection techniques on which our findings are based.

### 3.1 ClinMed approach

Within an initial phase of immersion hosted by Immanuel Clinic Rüdersdorf, health problems touching psychiatric and psychologic subjects were identified using several qualitative measures. Based on the principles of Design Thinking processes (Gavin Ambrose & Paul Harris, 2010), problems within the mental health care environment were defined by empathizing with the majority of stakeholders: service users, their representatives, and healthcare professionals, i.e. physicians, nurses and hospitals. Subsequently, all generated information were further analysed and elaborated using several tools. In our case, the *Customer Journey & Touch Points* instrument specifies the journey of a service user starting from the first thought of medical need to the final treatment. The journey, therefore, provides signs for touch points between service users and health care providers that are meant to be designed in a way that service user experience will be improved. Furthermore, the views of all stakeholders were analysed using the *Point-of-View* tool. To this end, individual stakeholders are characterized and subsequently related by lines and descriptions thereof. This reveals potential conflicts, problems and possible deducible future interactions.

The second phase of the summer school consisted of seven consecutive days, during which the project development was accompanied with lectures and coaching sessions. This phase addresses many aspects of the *Ideation & Prototyping* stages within the design process. Accompanied with knowledge transfer in the area of medical device development entailing e.g. regulatory, medico-economical and business topics, our team iteratively developed a solution to the previously elaborated challenges.

### 3.2 Data collection and analysis

Methodologically, we relied on qualitative data collection techniques: participant observation (n=3), focus group discussions (n=1) and informal conversations and interviews (n=2) to complement our observations. But, as this study reports from the results of a summer school, we also analysed notes from presentations (n=3), which often included discussions.

Data analysis followed an inductive approach and we applied open-coding. After identifying patterns, we categorized them. We used the open-source library for qualitative data analysis RQDA for R (Chandra & Shang, 2017). In a first step, we transcribed the data and created a new file in RQDA for each note, and assigned a file category such as "interview" or "observation". If necessary, German was translated into English by us. Next, we started with coding the files. These first-level codes were aggregated and categories were identified. Finally, we plotted these categories and exported the codes as html files to investigate and answer our research questions.

## 4 RESULTS

31 first-level codes were identified and four main categories were attributed to them. Figure 1 illustrates the result of the data analysis, with the categories highlighted. We found that some codes belong to more than one category. This is because the categories are closely related to each other. For example, the code "*afraid to use IT*" belongs to both "*communication*" and "*patient-centered care*". We see this on the one hand as a hinder for computer-supported communication, and on the other hand as a specific and relevant example how patient-centered care needs to be designed.

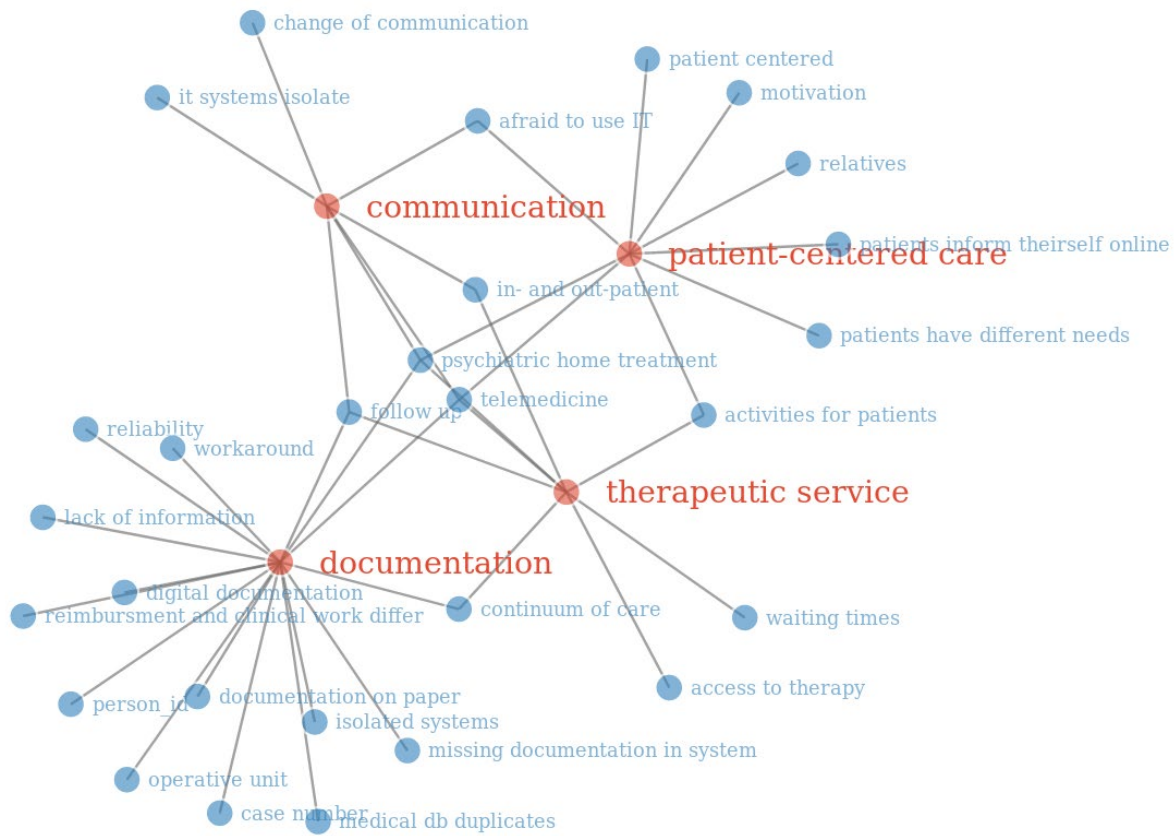

*Figure 1.    Notes were coded inductively, and aggregated to categories.*

The four main categories identified are "*communication*", "*documentation*", "*patient-centered care*" and "*therapeutic service*". These are illustrated in Figure 2, with some relevant excerpts for each category. In the following these categories are briefly introduced

- *Communication* refers to how service users communicate with their healthcare team, how health professionals communicate with each other, and how service users communicate with each other or their informal caregivers. During our fieldwork, the use of ICT to support communication was one of the focal points.

- The term "*documentation*" in our study refers to how care is documented, either on paper or electronically. We also included the comments on different systems, how they work together and support coordination of work.

- We use the Institute of Medicine's definition of *patient-centered care*: "providing care that is respectful of and responsive to individual patient preferences, needs, and values and ensuring that patient values guide all clinical decisions." (Committee on Quality of Health Care in America, 2001) In this category, comments and quotes that indicated a need or a pointed to an issue for patient-centered care were collected.

- The last category is "*therapeutic service*" and served as a pool for the different services provided by the psychiatric department, or services that were wished-for, either by the service users or healthcare professionals.

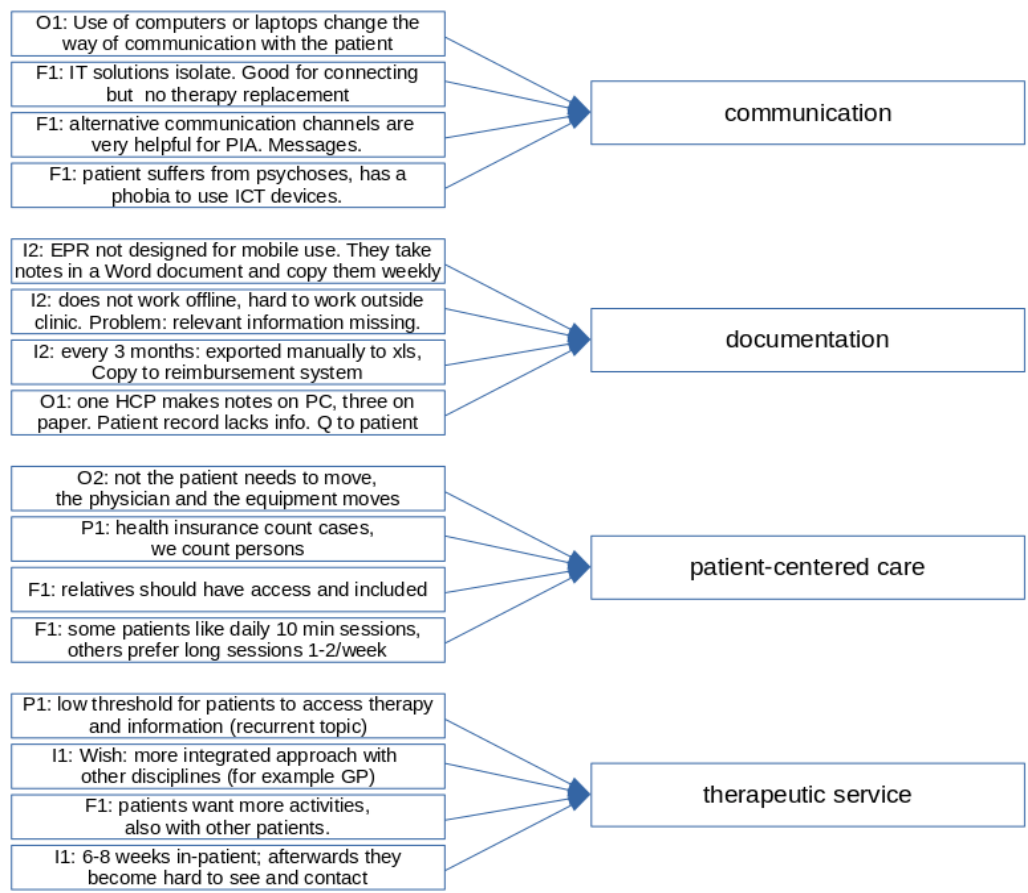

*Figure 2.    Significant quotes from the open coding at the first step, and the four main higher-level categories identified. (O...Observation, F...Focus group, I...Interview, P...Presentation)*

## 5    DISCUSSION

Psychiatric HT in Germany is in its very early stage. Since January 2018 first hospitals began to implement IET as a multi-professional and team-based approach to treat people with severe psychiatric issues in their home environment. It doesn't come as a surprise, that there are conflicts in this highly heterogeneous and complex context of mental health services. As IET is provided by hospitals it can be seen as a threat to local ambulatory clinics and psychotherapists seeing the hospital moving more into the outpatient sector (Bühring, 2018b) On the other hand, it can be seen as an alternative service from hospitals to reach the service users' specific needs, and also to reduce the costs of in-patient admissions (Bühring, 2018a).

We found that service users wish to have additional communication channels to stay in touch with their treatment team. Service users would welcome informal, asynchronous ways for non-critical communication. One service user used the metaphor of "*sympathetic ear*" ("*offenes Ohr*") to describe an app for communication and exchange of information. Also, healthcare professionals recognize the usefulness of additional communication channels for the psychiatric hospitals' outpatients department (PIA). On the other hand, both service users and healthcare professionals are sceptical towards introducing ICT solutions in their therapy ("*IT solutions are isolating. They are helpful to connect patients, but are not replacing therapy.*" - patient representative). Also, they refer to the available crisis line (*Telefonseelsorge*), which is accessible either by phone, mail or on-line chat. Another aspect of communication was brought to attention by one physician. He stated that an informal conversation after an observation, that "*the use of computers during therapy changes the nature of communication with the patient.*" In fact, this has been confirmed in an early study (Alsos et al., 2012).

"*Computers hinder work*". This quote of one the physicians sums up the overall feeling about documentation systems we have encountered. One of the main issues we have identified is that the documentation system used in HT was not designed for mobile use. The system does not support off-line use, which is the common mode of operation. HT takes often place in rural areas with bad or no reception. The healthcare professionals currently overcome this limitation by documenting the therapy locally in a Microsoft Word file. Once a week they enter the data in the documentation system. This is far from ideal, and can even lead to hazardous situations when for example a service user receiving home treatment is experiencing a psychiatric emergency and has to be admitted to hospital, while the latest, relevant treatment information is residing inaccessibly in a local document.

A similar workaround was reported to handle reimbursement processes. The psychiatric department uses mainly two systems for documentation: an Electronic Patient Record (EPR) system and an Enterprise Resource Planning (ERP) software. An interaction with a service user is registered in both systems manually, but to receive reimbursement for the HT from the health insurances, every three months these interactions are exported manually into a spreadsheet.

One of the clinic's key aims is to "*bridge in- and out-patient care*". The services provided by the psychiatric clinic are comprehensive: HT, in- and out- and (acute) day patient treatment - to name some outstanding examples. However, there is a lack of continuity between the different services offered, and also a lack of integration with other mainly outpatient health institutions. As one physician states: "*the patient is in-patient for 6-8 weeks, but afterwards it is hard to see him*". It was also noted several times, that continuity of care is important but it must ensure personnel continuity (Biringer et al., 2017), i.e. ensuring the same clinic team for the service users. ("*good outcome is only possible with good a relation to therapist*"). During a focus group session with service users, they stated that they lack variety and would like to have more activities coordinated, also between service users.

The department of psychiatry at the ICR is following a participative and collaborative approach both in practice and research. The clinic provides a low threshold access to therapy and information. By using different forms of therapy, the service users are "educated" - we could observe this during a group therapy session (reflecting team) where the service users can hear the physician reflecting over their case in front of them, and were invited to ask questions after the reflection. Further, the service users are organized and have a dedicated patient representative.

## 6 CONCLUSION

In this paper, we reported our insights from a two week summer school and picked an area which we believe is worth exploring in more detail: psychiatric HT.

We have described how IET as specific form of HT is designed in Germany, and how it is implemented at the specific context of the ICR. Our research questions focused on how HT is carried out, and how it is documented, coordinated and integrated. Several issues have been identified. Noteworthy is the issue of documentation, with a system that was not designed for this nature of work. Instead of supporting work, it is actually hindering it and, one might argue, is harmful in practice.

Interestingly, the idea of providing additional communication channels with the help of ICT was viewed controversially. There was a general consent, that it would "*be nice*" to have a complementary way of communication, especially for connecting service users or answering non-urgent questions. However, it is questionable if there is a strong benefit to have such systems.

The last research questions focused on patient participation. We have drawn a picture of what is currently in place. The design of new home treatment services can be seen as an opportunity, to engage the service user in the design process of such services. Although this might be challenging, with such a heterogeneous and sensitive group, we believe that service users' participation can increase the quality of the service and, naturally, the patient-centeredness.

We are well aware that the underlying data of this study is limited. This study should be seen as a first exploration towards the field of psychiatric home treatment, specifically of IET, from a CSCW perspective. We further believe that our findings are of interest for the study on welfare technology, even though our example of home treatment is rather low-tech, at least for now.

Future research is needed in the context of HT, especially for giving actual design recommendations for a system supporting the delivery of HT. This was the first paper of our newly formed collaboration, and we are planning to continue with a workshop in winter 2019/2020.

# ACKNOWLEDGEMENTS

The authors thank Karina Abdesselam, Ingmar van Hengel and Samzon Li for their contribution during the ClinMed 2019 summer school; the welcoming staff and patients at the Immanuel Clinic Rüdersdorf for providing insightful information; EIT Health for organising the summer school; the ClinMeD2019 team; the patient representative Peter Grollich for supporting our work; and Eric Monteiro for his valuable comments.